\documentclass[12pt]{article}
\usepackage{amsmath}
\usepackage{graphicx}
\begin{document}

\title{Self Gravitating Incompressible Fluid in Two Dimensions}
                   
\author{Mayer Humi and Zilu Tian\\
Department of Mathematical Sciences\\
Worcester Polytechnic Institute\\
100 Institute Road\\
Worcester, MA  01609}

\maketitle
\thispagestyle{empty}

\begin{abstract}
In this paper we develop two models for the steady states and evolution 
of two dimensional isothermal self gravitating and rotating incompressible 
gas which are based on the hydrodynamic equations for stratified fluid.
The first model is for the steady states of the gas while the
second addresses the time evolution of the gas subject to some
constraints. These models reduce the initial five partial differential
equations that govern this system to two for the steady state model and
to three for the time dependent model. Analytical and numerical solutions
of the model equations are used to study the structure of the resulting
steady and time dependent states of the fluid with some possible
astrophysical applications.
\end{abstract}

\newpage

\section{Introduction}

The steady states of self gravitating fluid in three dimensions have been 
studied by a long list of illustrious mathematical physicists. 
(For an extensive list of references see [1,2,3]). The motivation for
this research was due to the interest in the shape, stability and evolution
of celestial bodies and systems[12]. We now know however that many celestial 
objects such as galaxies and our solar system exhibit (effectively) 
"two dimensional structure" [4,5,6,7]. Furthermore recent discoveries are 
leading us to believe that systems similar to our solar system are 
"abundant" in the galaxy and their existence might be due to the collapse 
of a two dimensional interstellar cloud under gravitation (this is 
the so called the "nebular theory") [8,9,10,11,17]. This data leads us to 
believe that there is a fundamental physical process which we do not understand 
fully as yet that leads to the formation of planetary systems throughout 
the galaxy (and beyond).

This background motivates us to investigate in this paper the 
steady states and time-dependent evolution of a self gravitating 
and rotating fluid in two dimensions. This problem has 
been explored by a large number of investigators using elaborate analytic 
methods and computer simulations which involve, in general, thermodynamic 
considerations, magnetohydrodynamics modeling and turbulence.(For a complete 
list of references see [8,9,10,13,16]). While these are important issues
we still need, in our opinion, prototype analytic models that are able
to capture the evolution of this process and lead to insights about
its possible outcomes.

In this paper we attempt to develop such a model using the basic
hydrodynamic equations that govern the time-dependent evolution of an 
isothermal, incompressible, stratified  (i.e non constant density) 
and rotating fluid in two dimensions under gravity [1,2,3]. 
(The justification for the reduction from three to two dimensions
has been discussed by many authors. A lucid treatment is given in 
Ref. [12] pp.1-12).Under these assumptions we show that the number of model
equations can be reduced from five to a system of two equations for the
steady states and three coupled equations for its evolution. The models
contain some "parameter functions" which encode information about the
asymptotic mass density distribution of the fluid and its momentum.

The steady state model was investigated by us previously.[18,19,20]
However in this paper we consider a more general model
in which the "gas cloud" is rotating also with uniform angular 
velocity $\omega$ and study the impact of this rotation on the matter 
distribution in the steady state.

To study the predictions of these models we use both
analytical and numerical methods to solve their equations under a variety
of conditions. In particular we consider radial solutions to these equations
which represent the evolution of an interstellar cloud with
isothermal equation of state [10].

It might be argued that the hydrodynamic assumptions we are making in 
this paper are not realistic from astrophysical point of view. However 
our main goal is to capture analytically, as far as possible, the 
nonlinear and time dependent aspects of the processes under consideration. 
Accordingly our results might be useful to provide some analytic insights 
and guidelines for more elaborate work on this topic.

The plan of the paper is as follows: In Sec 2 we present the basic 
hydrodynamic equations and show how one can reduce them to a 
coupled system of three equations. Sec 3 presents further simplifications 
of these equations. The first is for the steady states of the model.
The second is for the time dependent evolution of the gas cloud under 
the assumption of constant vorticity. In Sec 4 we present analytical 
and numerical radial solutions of these equations. 
We end up Sec 5 with summary and conclusions.

\setcounter{equation}{0}
\section{Derivation of the Model Equations}

Following the standard convention  [1,14,15] we model the time dependent 
non-relativistic flow of an incompressible fluid in two dimensions $(x,y)$  
by the hydrodynamic equations of inviscid and incompressible stratified 
fluid 
\begin{equation}
\label{2.1}
u_x + v_y = 0
\end{equation}
\begin{equation}
\label{2.2}
\rho_t+u\rho_x + v\rho_y = 0
\end{equation}
\begin{equation}
\label{2.3}
\rho u_t+ \rho(uu_x+vu_y) = -p_x -\rho \phi_x +\rho \omega^2 x
\end{equation}
\begin{equation}
\label{2.4}
\rho v_t+\rho(uv_x+vv_y) = -p_y-\rho \phi_y +\rho \omega^2 y
\end{equation}
\begin{equation}
\label{2.5}
\nabla^2 \phi = 4 \pi G \rho
\end{equation}

where subscripts indicate differentiation with respect to the indicated
variable, ${\bf u}=(u,v)$ is the fluid velocity, $\rho$ is its density, 
$p$ is the pressure, $\phi$ is the gravitational field and G is 
the gravitational constant. The terms $\rho \omega^2 x$, $\rho \omega^2 y$
represent the components of the apparent centrifugal force due to the
rotation of the gas cloud with angular velocity $\omega$.

We can nondimensionalize these equations by introducing the following scalings
\begin{equation}
\label{2.6}
t=\frac{L{\tilde t}}{U_0},\,\,\, x= L\tilde{x},\,\,\, y=L\tilde{y},\,\,\, 
u=U_0 \tilde{u},\,\,\,v=U_0 \tilde{v},\,\,\, \rho = \rho_0 \tilde{\rho},\,\,\, 
p=\rho_0 U_0^2\tilde{p},\,\,\,\phi= U_0^2 \tilde{\phi},\,\,\,
\omega=\frac{U_0}{L}\tilde{\omega}.
\end{equation}
where $L,U_0,\rho_0$ are some characteristic length,velocity and mass density
respectively that characterize the problem at hand.
Substituting these scalings in eqs. (\ref{2.1})-(\ref{2.5}) and dropping 
the tildes these equations remain unchanged (but the quantities that appear
in these equations become nondimensional) while $G$ is replaced by 
$\tilde{G}=\frac{G\rho_0 L^2}{U_0^2}$. (Once again we drop the tilde).

In view of eq. (\ref{2.1}) we can introduce a stream function $\psi$ so that
\begin{equation}
\label{2.7}
u = \psi_y,\;\;v = -\psi_x\;.
\end{equation}

Using this stream function we can rewrite eq. (\ref{2.2}) as [13,15]
\begin{equation}
\label{2.8}
\rho_t+J\{\rho,\psi \}=0
\end{equation}
where for any two (smooth) functions $f,g$
\begin{equation}
\label{2.9}
J\{f,g\}=\frac{\partial f}{\partial x}\frac{\partial g}{\partial y} -
        \frac{\partial f}{\partial y}\frac{\partial g}{\partial x}
\end{equation}

Using $\psi$ the momentum equations (\ref{2.3}),(\ref{2.4}) become 
\begin{equation}
\label{2.10}
\rho(\psi_{yt}+\psi_y \psi_{yx} -\psi_x \psi_{yy}) = -p_x -\rho \phi_x
+\rho \omega^2 x
\end{equation}
\begin{equation}
\label{2.11}
\rho(-\psi_{xt}-\psi_y \psi_{xx} +\psi_x \psi_{xy}) = -p_y -\rho \phi_y
+\rho \omega^2 y
\end{equation}
To eliminate $p$ from these equations we differentiate eq. (\ref{2.10})
and eq. (\ref{2.11}) with respect to $y,x$ respectively and subtract.
This leads to
\begin{eqnarray}
\label{2.12}
&&\rho_y(\psi_{yt}+\psi_y \psi_{yx} -\psi_x \psi_{yy}) +
\rho(\psi_{yyt}+\psi_y \psi_{yyx} - \psi_x \psi_{yyy}) - \\ \notag 
&&\rho_x(-\psi_{xt}-\psi_y \psi_{xx} +\psi_x \psi_{xy}) - 
\rho(-\psi_{xxt}-\psi_y \psi_{xxx} +\psi_x \psi_{xxy}) = -J\{\phi,\rho\}+
J\{\frac{1}{2}\omega^2 r^2,\rho\}
\end{eqnarray}
where $r^2=x^2+y^2$.
The sum of the second and fourth terms in this equation can be rewritten  as
\begin{align}\label{2.13}
\rho(\nabla^2 \psi)_t+\rho J\{\nabla^2 \psi,\psi\}.
\end{align}

To reduce the first and third terms in (\ref{2.12}) we use (\ref{2.8}).
We obtain
\begin{align}\label{2.14}
\rho_y(\psi_{yt}+\psi_y \psi_{yx} -\psi_x \psi_{yy})-
\rho_x(-\psi_{xt}-\psi_y \psi_{xx} +\psi_x \psi_{xy})= \\ \notag
\rho_y(\psi_{yt}+\rho_y\psi_y\psi_{yx}-(\rho_t+\rho_x\psi_y)\psi_{yy}
+\rho_x\psi_{xt}+(\psi_x\rho_y-\rho_t)\psi_{xx}-\rho_x\psi_x\psi_{xy}= \\ \notag
\rho_y\psi_{yt}+\rho_y\psi_{yt}-\rho_t\nabla^2\psi+
\frac{1}{2}J\{(\psi_x)^2+(\psi_y)^2,\rho\}.
\end{align}
combining the results of (\ref{2.13}) and (\ref{2.14}) eq. (\ref{2.12}) becomes
\begin{align}\label{2.15}
\rho_y\psi_{yt}+\rho_x\psi_{xt} -\rho_t\nabla^2\psi + \rho(\nabla^2 \psi)_t+
\rho J\{\nabla^2 \psi,\psi\}+\frac{1}{2}J\{(\psi_x)^2+(\psi_y)^2,\rho\} \\ \notag
=-J\{\phi,\rho\}+J\{\frac{1}{2}\omega^2 r^2,\rho\}
\end{align}
Thus we have reduced the original five equations (\ref{2.1})-(\ref{2.5})
to three equations (\ref{2.5}), (\ref{2.8}) and (\ref{2.15}). Although
(\ref{2.15}) is rather cumbersome in general, it can simplified further
under some restrictions which are presented in the following section.

\setcounter{equation}{0}
\section{Simplification of the Model Equations}

Equation (\ref{2.15}) can be simplified further in two cases.
The first is when we consider only steady states 
of the flow and the second is when the flow vorticity is constant.

\subsection{A Model for the Steady States}

When we consider only steady states of the flow (\ref{2.8}) implies that 
$\psi=\psi(\rho)$ and after some algebra [18] (\ref{2.15}) reduces to 
\begin{equation}
\label{4.1}
H(\rho)^{1/2}\nabla {\bf{\cdot}} ( H(\rho)^{1/2}\nabla \rho)  
+\phi-\frac{1}{2}\omega^2 r^2 =S(\rho).
\end{equation}
Where 
\begin{equation}
\label{4.1a}
H(\rho)=\rho\psi_{\rho}^2
\end{equation}
and $S(\rho)$ is some function of $\rho$. 
Thus the equations governing the steady state are (\ref{4.1}), (\ref{2.5})
and $H(\rho)$ and $S(\rho)$ are "parameter functions" which determine 
the nature of the steady state.

\subsubsection{The Physical Meaning of the Functions $H(\rho)$, $S(\rho)$}

The function $H(\rho)$ is a
parameter function which is determined by the momentum (and angular
momentum) distribution in the fluid. From a practical point of view the
choice of this function determines the structure of the steady state
density distribution. The corresponding flow field can be computed then
aposteriori (that is after solving for $\rho$) from the following 
relations.[18]
\begin{equation}
\label{3.1}
u=\sqrt{\frac{H(\rho)}{\rho}} \frac{\partial \rho}{\partial y}, \,\,\,
v=-\sqrt{\frac{H(\rho)}{\rho}} \frac{\partial \rho}{\partial x}.
\end{equation}

The function $S(\rho)$ that appears in eq. (\ref{4.1}) can be determined 
from the asymptotic values of $\rho$ and $\phi$ on the boundaries of the
domain on which eqs (\ref{2.5}),(\ref{4.1}) are solved. 
When these asymptotic values are imposed or known one can
evaluate the left hand side of eq. (\ref{4.1}) on the domain boundaries
and re-express it in terms of $\rho$ only to determine $S(\rho)$
on the boundary of the domain. However, the resulting functional relationship 
of $S$ on $\rho$ must then hold also within the domain itself since $S$ 
does not depend on x, y directly. 

For example if we assume that on an infinite
domain $h(\rho)=1$, $\omega=0$ and the asymptotic behavior of $\rho$ 
and $\phi$ is given by
\begin{equation}
\label{3.2}
\displaystyle\lim_{r\rightarrow \infty} \rho(r)=e^{-\alpha r^2},\,\,\,
\displaystyle\lim_{r\rightarrow \infty} \phi(r)=4\alpha^2 r^2 e^{-\alpha r^2}
\end{equation}
then (asymptotically) (\ref{4.1}) evaluates to  
\begin{equation}
\label{3.3}
S(\rho)=-4\alpha e^{-\alpha r^2} = -4\alpha\rho
\end{equation}

\subsection{A model for the Time Evolution}

To begin with we consider the case where the vorticity is zero and
then generalize to the case where the flow vorticity is constant.
 
When the flow vorticity $\nabla \times  {\bf u}$ is zero 
then $\nabla^2 \psi=0$ eq. (\ref{2.15}) becomes
\begin{align}\label{2.16}
\rho_y\psi_{yt}+\rho_x\psi_{xt}+\frac{1}{2}J\{(\psi_x)^2+(\psi_y)^2,\rho\}
=-J\{\phi,\rho\} 
\end{align}
However when the vorticity is zero we can introduce the velocity potential 
$\eta$ which satisfies $\eta_x=u$, $\eta_y=v$. Replacing $\psi$ by $\eta$
in (\ref{2.16}) we obtain
\begin{equation}
\label{2.17}
J\{\eta_t+\frac{1}{2}[(\eta_x)^2+(\eta_y)^2]+\phi,\rho\} =0
\end{equation}
Hence
\begin{equation}
\label{2.18}
\eta_t+\frac{1}{2}[(\eta_x)^2+(\eta_y)^2]+\phi =S(\rho)
\end{equation}
The equations of the flow in this case are 
\begin{equation}
\label{2.19}
\rho_t+\eta_x\rho_x+\eta_y\rho_y=0
\end{equation}
(which replaces (\ref{2.8})), (\ref{2.18}) and (\ref{2.5}).

To generalize this reduction to the case where $\nabla^2 \psi =a$ 
(where $a$ is any constant) we define
$$
v_1=\psi_y,\,\,\, v_2=-\psi_x+ax
$$
Therefore
$$
(v_1)_y-(v_2)_x=0,
$$
which implies that there exists a function $\eta$ so that
$$
\eta_x=v_1,\,\,\, \eta_y=v_2.
$$
Hence
\begin{align}
\label{2.20}
\eta_x=\psi_y,\,\,\,  \eta_y=-\psi_x+ax
\end{align}
Using these relations to substitute $\eta$ for $\psi$ in (\ref{2.23})
leads to
\begin{align}\label{2.22}
\rho_y\eta_{xt}-\rho_x(\eta_y -ax)_t+
\left[-a\rho_t+\frac{1}{2}J\{(\eta_y-ax)^2+(\eta_x)^2,\rho\}\right]
=-J\{\phi,\rho\}.
\end{align}
Therefore
\begin{align}\label{2.23}
J\{\eta_t,\rho\}
-a\rho_t+\frac{1}{2}J\{(\eta_y-ax)^2+(\eta_x)^2,\rho\}
=-J\{\phi,\rho\}.
\end{align}
Hence 
\begin{equation}
\label{2.24}
-a\rho_t+
J\{\eta_t+\frac{1}{2}[(\eta_y-ax)^2+(\eta_x)^2]+\phi,\rho\}=0.
\end{equation}
Using (\ref{2.8}) we have
\begin{equation}
\label{2.25}
-aJ\{\psi,\rho\}+
J\{\eta_t+\frac{1}{2}[(\eta_y-ax)^2+(\eta_x)^2]+\phi,\rho\}
\end{equation}
It follows then that
\begin{equation}
\label{2.26}
-a\psi+ \eta_t+\frac{1}{2}[(\eta_y-ax)^2+(\eta_x)^2]+\phi = S(\rho).
\end{equation}
If $a \ne 0$, $\psi$ can be eliminated from this equation if we differentiate 
with respect to $y$ and use (\ref{2.20}) to obtain
\begin{equation}
\label{2.27}
-a\eta_x+ \left[\eta_t+\frac{1}{2}[(\eta_y-ax)^2+(\eta_x)^2]+\phi\right]_y
=S(\rho)_y
\end{equation}

\setcounter{equation}{0}
\section{Radial Solutions for the Steady State Model}

When we consider the special case where in polar coordinates $\rho=\rho(r)$
and $\phi=\phi(r)$ the system (\ref{4.1}) and (\ref{2.5}) with $H(\rho)=1$
reduces to 
\begin{equation}
\label{4.2}
\rho^{\prime\prime}=-\frac{\rho^{\prime}}{r}+S(\rho) -\phi +
\frac{1}{2}\omega^2 r^2
\end{equation}
\begin{equation}
\label{4.3}
\phi^{\prime\prime}=-\frac{1}{r}\phi^{\prime}+c\rho,\,\,\, c=4\pi G
\end{equation}
To solve this system of equations we let $S(\rho)=\alpha\rho$, solve 
(\ref{4.2}) for $\phi$ and substitute the result in (\ref{4.3}). This leads 
to the following fourth order equation for $\rho$
\begin{equation}
\label{4.4}
\rho''''+\frac{2}{r}\rho'''-\left(\alpha+\frac{1}{r^2}\right)\rho''+
\left(\frac{1}{r^3}-\frac{1}{r}\right)\rho'+c\rho=2\omega^2.
\end{equation}
The general solution of this equation is
\begin{equation}
\label{4.5}
\rho=\frac{2\omega^2}{c}+C_1J_0(a_1r)+C_2J_0(b_1r)+C_3Y_0(a_1r)+C_4Y_0(b_1r)
\end{equation}
where $J_0$ and $Y_0$ are Bessel functions of the first and second kind of 
order $0$ and 
$$
a_1=\frac{1}{2}\sqrt{-2\alpha+2\sqrt{\alpha^2-4c+\alpha^2}},\,\,\,
b_1=\frac{1}{2}\sqrt{-2\alpha-2\sqrt{\alpha^2-4c+\alpha^2}}
$$
Assuming no singularity at the origin we set $C_3=C_4=0$.
To assess the impact of the rotation term on the steady state we solved
this system for $C_1,\,C_2$ on a circular disk using the boundary 
conditions $\rho(0)=1$ and $\rho(8)=0$ with $c=1$, $\alpha=-19.4$.
The results of these computations for different values of $\omega$ are 
plotted in Fig. $1$. In this figure we see that the separation between the 
density peaks become more pronounced as $\omega$ increases. This might 
interpreted as leading to the creation of protoplanets around the central 
core.

A strong dependence on $\omega$ is shown in 
Fig. $2$ which has the same parameters as Fig. $1$ except that the
boundary conditions on $\rho$ are: $\rho(0)=0.35$ and $\rho(8)=0.25$.
This figure illustrate clearly the effect that rotation can have on
the pattern of density fluctuations within the cloud. Furthermore
in this figure the magnitude of the density fluctuations reverses itself as
$\omega$ becomes larger  viz. the higher density peaks are placed
at larger values of $r$. (Which is reminiscent of the situation in 
the solar system)

\section{Radial Solutions for the Time Evolution Model}
The system (\ref{2.5}),(\ref{2.18}) and (\ref{2.19}) can be simplified
further if we use polar coordinates and assume
that $\rho,\,\eta,\,\phi$ are functions of $r$ and $t$ only. We obtain,
\begin{align}
\label{2.28}
\rho_t+\eta_r\rho_r=0, \\ \notag
\phi_{rr}+\frac{1}{r}\phi_r-c\rho=0, \\ \notag
\eta_t+\frac{1}{2}(\eta_r)^2+\phi=S(\rho).
\end{align}
where $c=4\pi G$.

\subsection{Steady States}

When we consider a steady state solutions of (\ref{2.28}) then $\rho_t=0$
and $\eta_t=0$. If follows from the first equation in (\ref{2.28})
that either $\rho_r$ or $\eta_r$ must be zero. In the first 
case $\rho$ is constant and we can let $\rho=1$ without loss of generality.
When $\eta_r$ is zero we must have $\phi=S(\rho)$ and the second equation
in (\ref{2.28}) becomes
\begin{align}
\label{2.29}
S'(\rho)\left[\rho_{rr}+\frac{1}{r}\rho_r\right]+S''(\rho)(\rho_r)^2-
c\rho = 0
\end{align}
where primes denote differentiation with respect to $\rho$.

We consider these two cases separately.

A. Steady state with $\rho=1$

Since $\rho=1$ the function $S(\rho)$ is a constant and 
the general solution for $\phi$ is 
\begin{equation}
\label{2.30}
\phi=\frac{c}{4}r^2 + C_1\ln r + C_2.
\end{equation}
where $C_1$, $C_2$ are arbitrary constants. The equation for $\eta$ becomes
\begin{equation}
\label{2.31}
\frac{1}{2}(\eta_r)^2=S-\frac{c}{4}r^2-C_1\ln(r)- C_2.
\end{equation}
($S$ can be absorbed in $C_2$ but we leave it in this form as these two 
constants have different physical meaning). If we let $C_1=0$ to avoid the 
singularity at the origin (\ref{2.31}) yields
\begin{equation}
\label{2.32}
\eta=\pm\left\{\frac{1}{4}r\sqrt{8S-8C_2-2cr^2}+\frac{(S-C_2)\sqrt{2}}
{\sqrt{c}}\arctan\left[\frac{\sqrt{2c}\,r}{\sqrt{8S-8C_2-2cr^2}}\right]\right\}
+C_3
\end{equation}

B. Steady states with $\eta=1$

In this case the solution of (\ref{2.29}) depends on the nature of the 
function $S(\rho)$. In general this equation has to be solved numerically.
However we present here analytical solutions of this equation for two
special cases.
\begin{enumerate}

\item $S(\rho)=\alpha\rho$ where $\alpha$ is a constant.
The solution to (\ref{2.29}) in this case is
\begin{equation}
\label{2.33}
\rho=C_4J_0\left(\sqrt{-\frac{c}{\alpha}}\,r\right)+
C_5Y_0\left(\sqrt{-\frac{c}{\alpha}}\,r\right)
\end{equation}
It follows then that the nature of the steady state is determined by the 
ratio $\frac{c}{\alpha}$. A sample of the resulting $\rho$ profiles is
presented in $Fig.\,3$. To obtain this figure we considered a pinched disk 
with $\rho(0.01)=1$, $\rho'(0.01)=-10$ and $c=\alpha$. 
The resulting steady state has an increase in the material density towards the 
circumference of the disk. Similar graphs were obtained numerically for 
$S(\rho)=\alpha\rho^n$, $n=2,3$.

\item $S(\rho)=\alpha\ln(\rho)$

In this case we have
\begin{equation}
\label{2.34}
\rho=\frac{1}{2C_1cr^2\cos(\theta)^2}
\end{equation}
where
$$
\theta=\frac{1}{2\sqrt{C_1\alpha}}(\ln r-C_2)
$$
Substituting $C_1=C_2=c=\alpha=1$ we obtain $Fig.\,4$ which might be 
interpreted as representing a binary system.
\end{enumerate}
 
\subsection{Perturbations from the steady state $\rho=1$}

We consider in this section a disk of radius $1$ with a steady state
$\rho_0=1$ and $S(\rho)=0$. Letting $\phi(1)=0$ and using (\ref{2.30}), 
(\ref{2.32}) (with $C_1=0$) this yields the following equations for the 
steady state
$$
\phi_0(r)=\frac{c}{4}(r^2-1) 
$$
$$
\eta_0=\frac{\sqrt{2c}}{4}\left[\arcsin(r)+r\sqrt(1-r^2)\right]
$$

For a perturbation from this state, viz.
\begin{equation}
\label{2.35}
\rho(t,r)=\rho_0+\epsilon \rho_1(t,r),\,\,\, 
\phi(t,r)=\phi_0+\epsilon \phi_1(t,r),\,\,\,
\eta(t,r)=\eta_0+\epsilon \eta_1(t,r)
\end{equation}
we obtain to first order in $\epsilon$ the following system of equations:
\begin{align}
\label{2.36}
(\rho_1)_t+\frac{\sqrt{2c(1-r^2)}}{2}(\rho_1)_r=0, \\ \notag
(\phi_1)_{rr}+\frac{1}{r}(\phi_1)_r-c\rho_1=0, \\ \notag
(\eta_1)_t+\frac{\sqrt{2c(1-r^2)}}{2}(\eta_1)_r+\phi_1=0.
\end{align}
The equation for $\rho_1$ in (\ref{2.36}) can be solved analytically. 
Its general solution is
\begin{equation}
\label{2.37}
\rho_1=F\left(\sqrt{\frac{2}{c}}\arcsin(r)-t\right)
\end{equation}
where $F$ is any smooth function of its variable which has to be adjusted to 
the initial conditions of the perturbation.
The second equation in (\ref{2.36}) is a (reduced) Poisson equation and 
its general solution can be expressed by quadratures
\begin{equation}
\label{2.37a}
\phi_1=\int\frac{c\{\int r F\left(\sqrt{\frac{2}{c}}\arcsin(r)-t\right)
\,dr+F_1(t)\}}{r}\,dr+F_2(t)
\end{equation}
where $F_1(t)$, $F_2(t)$ have to be determined by the boundary 
conditions on $\phi_1$. Finally one can obtain also an expression 
for the solution for $\eta$ in terms of quadratures.

For example if the initial perturbation in $\rho$ is $\rho_1(0,r)=ar$ 
where $a$ is a constant then
$$
F(x)=a\sin\left(\sqrt{\frac{c}{2}}x\right)
$$
and 
\begin{equation}
\label{2.37b}
\rho_1(t,r)=a\sin\left(\arcsin(r)-\sqrt{\frac{c}{2}}t\right)=
a\left\{r\cos\left(\sqrt{\frac{c}{2}}\,t\right)-
\sqrt{1-r^2}\sin\left(\sqrt{\frac{c}{2}}\,t\right)\right\}.
\end{equation}
The evaluation of $\phi_1$ using the second equation in (\ref{2.36}) and
(\ref{2.37b}) is straightforward. 
It should be obvious how one can generalize this example to other
expressions for $\rho_1(0,r)$.

A second approach to the solution of the system (\ref{2.36}) is to assume 
exponential dependence in time, viz.
\begin{equation}
\label{2.38}
\rho_1=e^{\alpha t} R(r),\,\,\,\eta_1=e^{\alpha t} E(r),\,\,\,
\phi_1=e^{\alpha t} P(r)
\end{equation}
This ansatz reduces (\ref{2.36}) to a system of ordinary differential
equations
\begin{align}
\label{2.39}
\frac{\sqrt{2c(1-r^2)}}{2}R(r)'+\alpha R(r)=0, \\ \notag
P(r)''+rP(r)'-crR(r)=0, \\ \notag
\frac{\sqrt{2c(1-r^2)}}{2}E(r)'+\alpha E(r) +P(r)=0.
\end{align}
As before the equation for $R(r)$ can be solved analytically,
$$
R(r)=C_1\exp\left(-\alpha\sqrt{\frac{2}{c}}\arcsin(r)\right)
$$
while the equations for $P(r)$ and $E(r)$ can be solved by quadratures
or numerically.

A numerical approach to the solution for $\phi_1$ and $\eta_1$ in
(\ref{2.36}) is also possible.

%We simulated numerically the solution of (\ref{2.36}) for $c=0.01$
%with an initial perturbation to the steady state in the form
%$\rho_1(0,r)=0.1 \sin(4\pi r)$.

%The evolution of this perturbation was considered for two situations. 
%The first without the presence of a protostar at the center of the disk 
%(Fig. $3$) and the other with it (Fig. $4$). In each of these figures 
%we present the state of the perturbation at $t=0.1,\,0.4,\,0.7$ and $t=1$.  
%In all cases the error limit for each (adaptive) time step in the simulation
%was set to $10^{-5}$.

%Figure $3$ shows that in the absence of a protostar the perturbation peak 
%is moving forward in time. In the presence of a protostar (Fig. $4$)
%however the matter density near the protostar increases as matter is being
%"sucked" into it. 

\subsection{Perturbations from the steady state $\eta_0=1$}

As in the previous subsection we consider again a disk of radius $1$
and let $S(\rho)=\alpha\rho$. The general steady state solution for $\rho$ 
is given by (\ref{2.32}). Assuming no singularities in $\rho$ (ie. no 
protostar at the origin) we must set $C_2=0$ in this equation. Furthermore
since $\rho \ge 0$ it follows that we must have $\sqrt{-c/\alpha}=\beta$
where $\beta$ is the first zero of $J_0$. Thus
$$
\rho_0=J_0(\beta r),\,\,\,\phi_0=\alpha\rho_0
$$
(where we normalized $rho_0$ at $r=0$ to be $1$).

For a perturbation from this steady state in the form given by (\ref{2.35})
we obtain to first order in $\epsilon$ the following system of equations:
\begin{eqnarray}
\label{2.40}
&&(\phi_1)_{rr}+\frac{1}{r}(\phi_1)_r-c\rho_1=0, \\ \notag
&&(\eta_1)_t-\alpha\rho_1+\phi_1=0 \\ \notag
&&(\rho_1)_t-\beta J_1(\beta r)(\eta_1)_r=0
\end{eqnarray}
Where $J_1$ is Bessel function of  the first kind of order $1$.
The evolution of an initial perturbation $\rho_1=exp(-5r)$ with 
$\alpha=0.01$ from the steady state is plotted in $Fig 5$. This figure 
shows that as time progresses there is an accumulation of matter near the 
center of the disk. At the same time there is an initial separation 
between the core and the rest of the disk. 

We computed also the solution to the system (\ref{2.28}) with an initial
matter distribution $\rho(0,r)=\frac{1+\sin(2\pi r)}{2}$, $c=0.01$ and
$t \in [0,9]$. The results of the  simulation (Fig. $6$) show that as 
time progresses matter is starting to build up in the vicinity
of the center of the disk and around $r=1$. At the same time there is 
a decrease in matter density in between these two points.

\setcounter{equation}{0}
\section{Summary and Conclusions}

In previous publications [18-20] we treated only the steady states of two 
dimensional self gravitating fluid. In this paper we generalized this model 
to include disk rotation and assessed the impact of this addition on the 
distribution of matter in the disk. We were able also to 
address the time dependent evolution of this fluid under restrictions on 
its vorticity. This enabled us to simplify considerably the equations 
which govern its evolution. While this is a highly idealized model in 
the context of astrophysical applications it may still provide some 
analytical insights for more elaborate models. 

In this paper we considered only radial solutions of this model.
More general solutions which are not radial will have to be explored next.

\newpage
\centerline{\Large{\bf List of Captions}}

\vspace*{.50in}

\begin{tabular}{ll}

Fig. 1 Steady states with $\alpha=-19.4$, $c=1$ and boudary conditions
$\rho(0)=1$,\\  $\rho(8)=0$ with different values of $\omega$
\\
Fig. 2 Steady states with $\alpha=-19.4$, $c=1$ and boudary conditions
$\rho(0)=0.35$,\\ $\rho(8)=0.25$ with different values of $\omega$
\\
Fig. 3 The steady state that corresponds to (\ref{2.33})
\\
Fig. 4 The steady state that corresponds to (\ref{2.34})
%Fig. 1  Using (\ref{2.30}) to solve for $\rho_1$ with $c=0.01$, \\
%and initial perturbation $\rho_1=0.1e^{-r}$. No protostar at the origin.
%\\
%Fig. 3  Using (\ref{2.36}) to solve for $\rho_1$ with $c=0.01$, \\
%initial perturbation $\rho_1=0.1\sin(4\pi r)$. No protostar at the origin.
%\\
%Fig. 3  Using (\ref{2.36}) to solve for $\rho_1$ with $c=0.01$, \\
%and initial perturbation $\rho_1=0.1e^{-r}$ with a protostar at the origin.
%\\
%Fig. 4  Using (\ref{2.36}) to solve for $\rho_1$ with $c=0.01$, \\
%initial perturbation $\rho_1=0.1\sin(4\pi r)$ with a protostar at the origin.
\\
Fig. 5  Using (\ref{2.40}) to solve for $\rho_1$ with $\alpha=-0.01$, \\
$c=0.0578$ and
initial perturbation $\rho_1=exp(-5r)$ with a protostar at the origin.
\\
Fig. 6  Using (\ref{2.28}) to compute the evolution of $\rho$. The initial 
matter distribution is \\
$\rho(0,r)=\frac{1+\sin(2\pi r)}{2}$ and $c=0.01$. No protostar at the origin.
\end{tabular}

\newpage
\begin{figure}[ht]
\centerline{\includegraphics[height=100mm,width=120mm,clip,keepaspectratio]{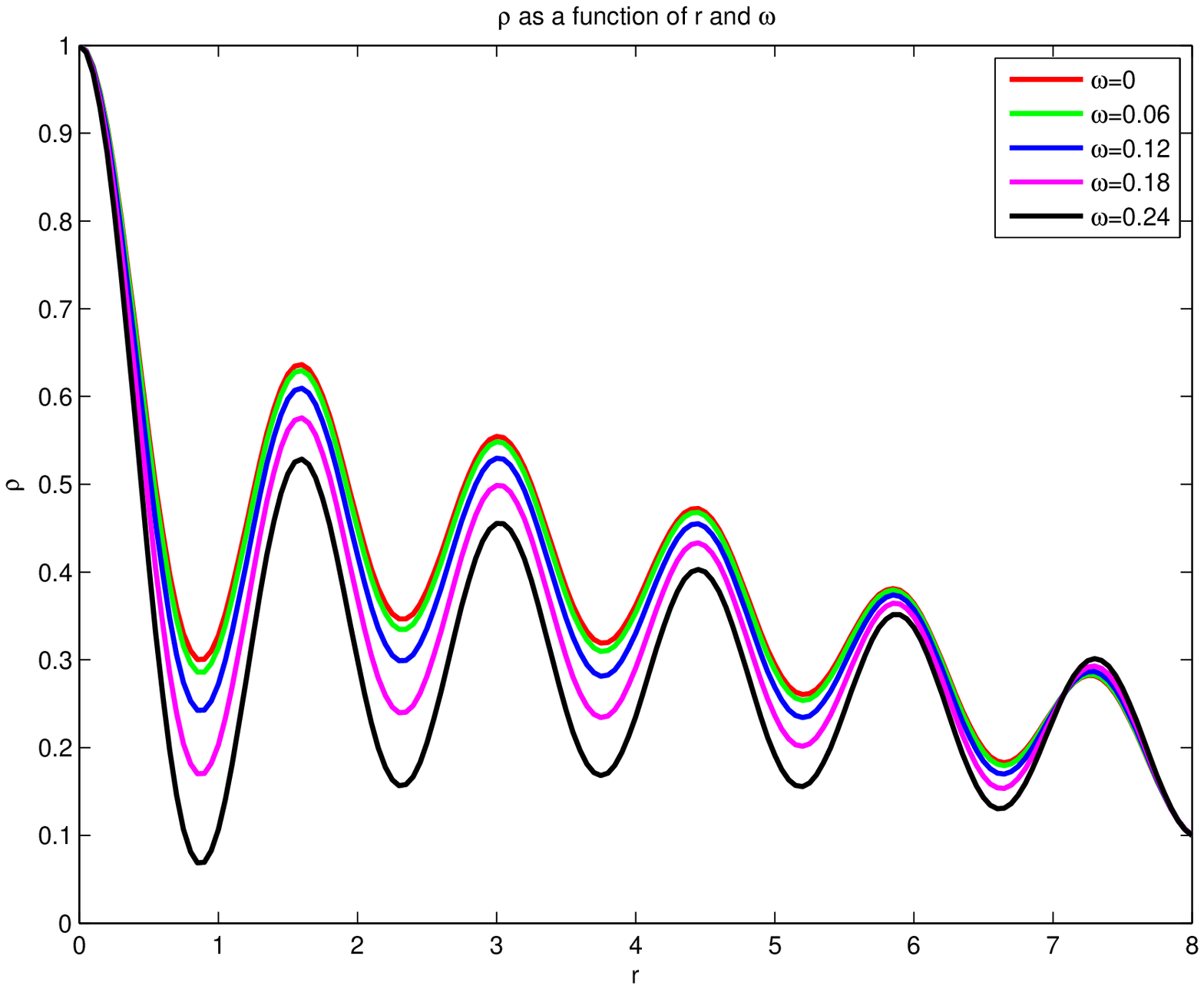}}
\label{Figure 1}
\caption{}
\end{figure}

\newpage
\begin{figure}[ht]
\centerline{\includegraphics[height=100mm,width=120mm,clip,keepaspectratio]{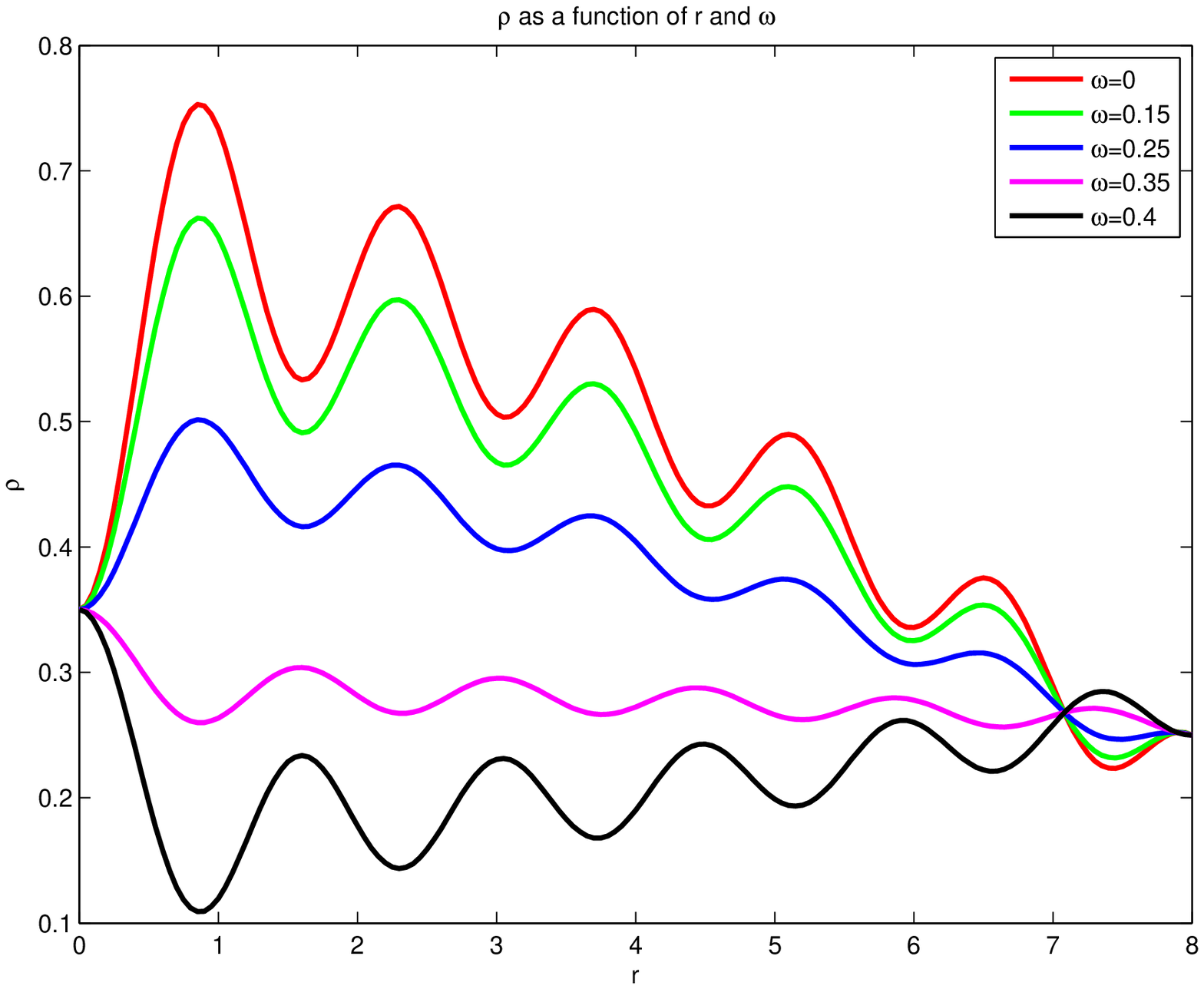}}
\label{Figure 2}
\caption{}
\end{figure}

\newpage
\newpage
\newpage
\begin{figure}[ht]
\centerline{\includegraphics[height=100mm,width=120mm,clip,keepaspectratio]{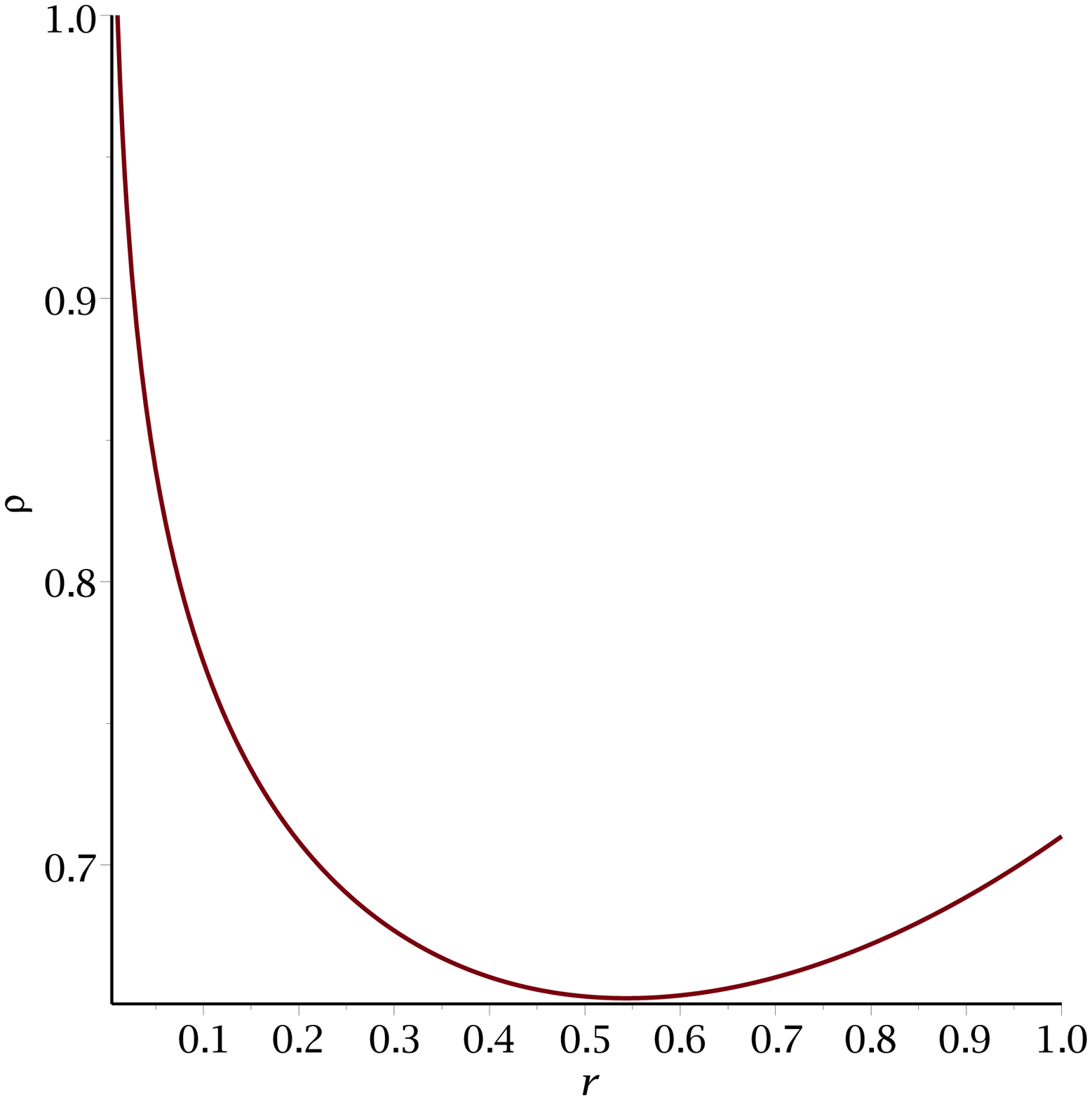}}
\label{Figure 3}
\caption{}
\end{figure}

\newpage
\begin{figure}[ht]
\centerline{\includegraphics[height=100mm,width=120mm,clip,keepaspectratio]{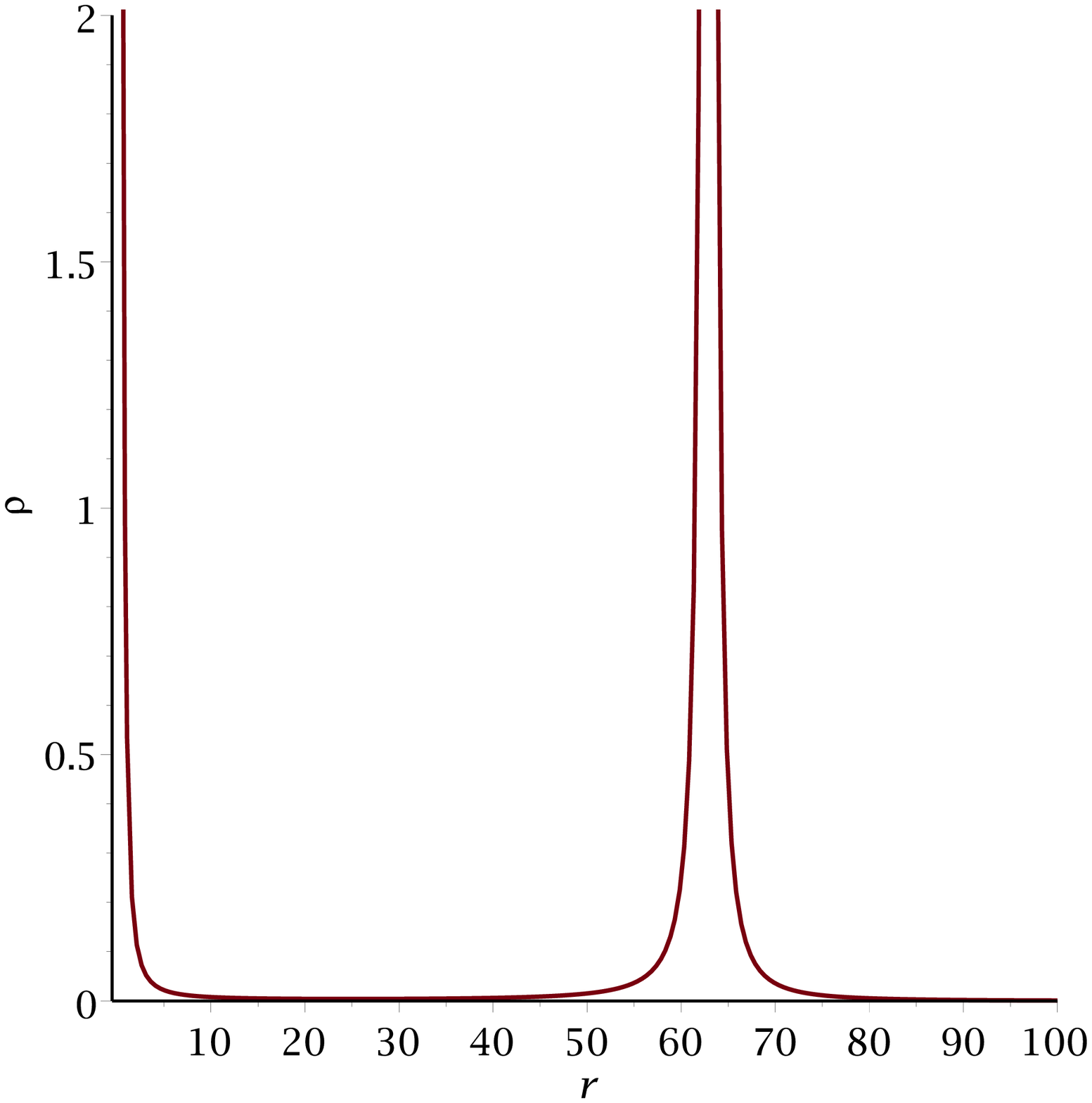}}
\label{Figure 4}
\caption{}
\end{figure}

%\newpage
%\newpage
%\begin{figure}[ht]
%\centerline{\includegraphics[height=100mm,width=120mm,clip,keepaspectratio]{fig02.ps}}
%\label{Figure 3}
%\caption{}
%\end{figure}

%\newpage
%\begin{figure}[ht]
%\centerline{\includegraphics[height=100mm,width=120mm,clip,keepaspectratio]{fig03.ps}}
%\label{Figure 3}
%\caption{}
%\end{figure}

%\newpage
%\begin{figure}[ht]
%\centerline{\includegraphics[height=100mm,width=120mm,clip,keepaspectratio]{fig04.ps}}
%\label{Figure 4}
%\caption{}
%\end{figure}

\newpage
\begin{figure}[ht]
\centerline{\includegraphics[height=100mm,width=120mm,clip,keepaspectratio]{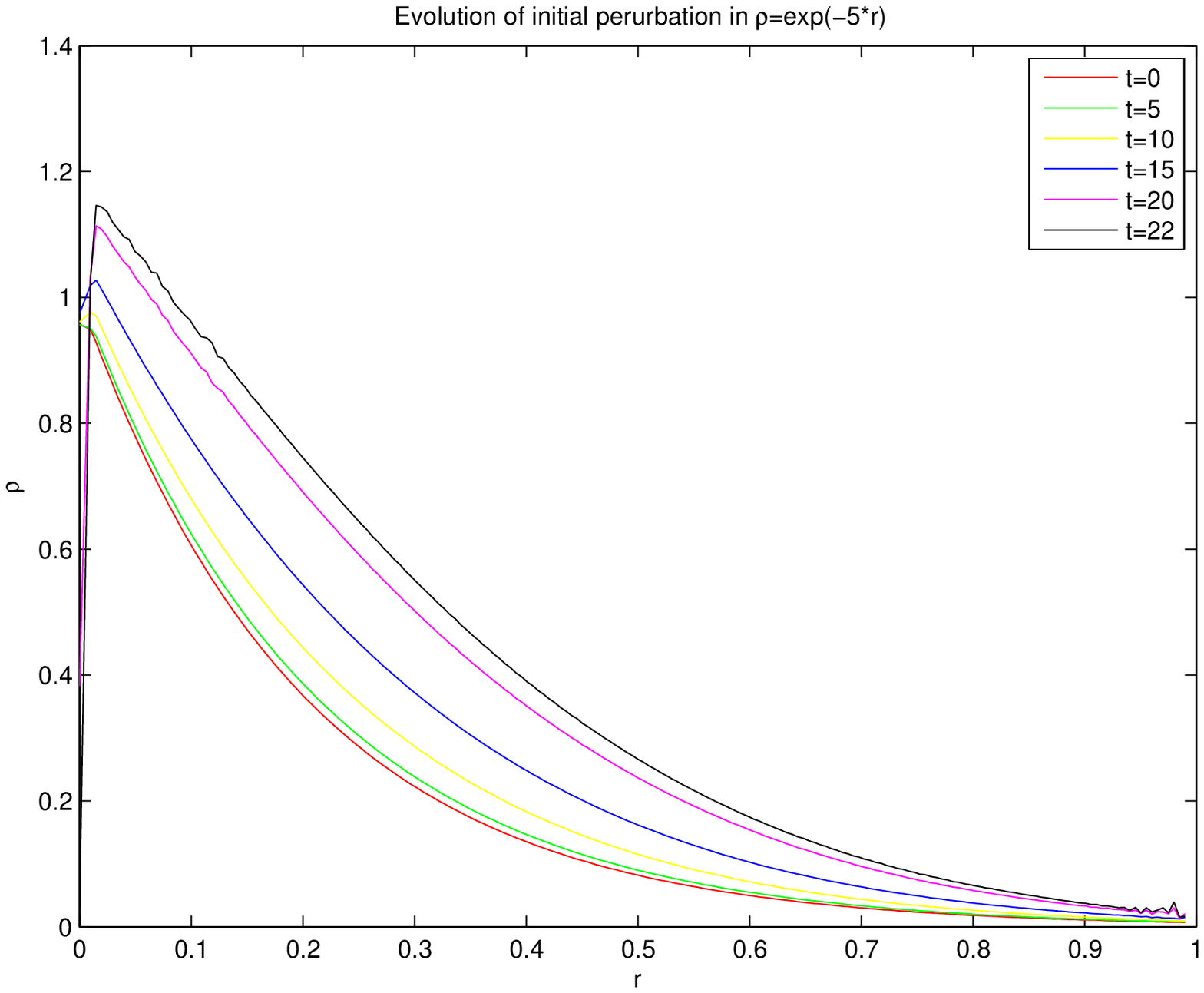}}
\label{Figure 5}
\caption{}
\end{figure}

\newpage
\newpage
\begin{figure}[ht]
\centerline{\includegraphics[height=100mm,width=120mm,clip,keepaspectratio]{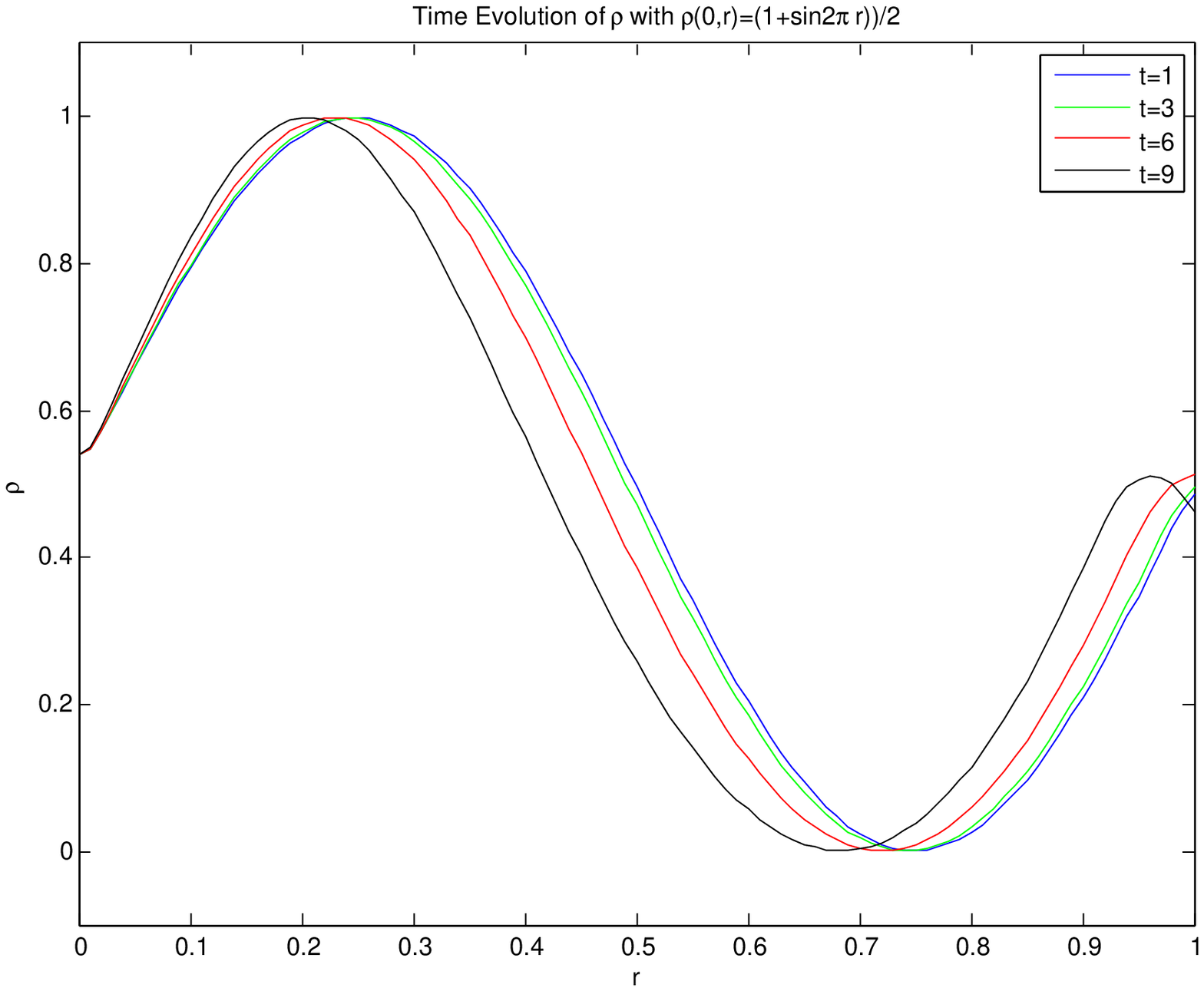}}
\label{Figure 6}
\caption{}
\end{figure}

\end{document}